\documentclass{nature}
\usepackage{amsmath,mathrsfs,amsbsy,color,graphicx,bm,amsthm,amsfonts}
\usepackage{units}
\usepackage{bbm}
\usepackage{times}
\usepackage{dcolumn}
\usepackage{amsmath,amssymb,epsfig,float}

\begin{document}
\baselineskip=0.8 cm
\title{Strong Coherent Light Amplification with Double Electromagnetically Induced Transparency Coherences}
\author{Dan Wang$^{1,2,3}$, Chao Liu$^{2,3}$, Changshun Xiao$^{2,3}$,
Junxiang Zhang$^{1,3\ast}$, Hessa M. M. Alotaibi$^{4}$, Barry C. Sanders$^{5,6,7,8 }$, Li-Gang Wang$^{1}$, \& Shiyao Zhu$^{1,9}$}
\maketitle

\begin{affiliations}
\item
Department of Physics, Zhejiang University, Hangzhou 310027, China.
$^{2}$ College of Physics and Electronic Engineering, Shanxi University, Taiyuan 030006, China. $^{3}$ Collaborative Innovation Center of Extreme Optics, Shanxi University, Taiyuan 030006, China. $^{4}$ Public Authority for Applied Education and Training, P.O. Box 23167, Safat 13092, Kuwait.
$^{5}$ Hefei National Laboratory for Physical Sciences at Microscale,
University of Science and Technology of China, Hefei 230026, China.
$^{6}$ Shanghai Branch, CAS Center for Excellence and Synergetic Innovation Center in Quantum Information and Quantum Physics, University of Science and Technology of China, Shanghai 201315, China.
$^{7}$ Institute for Quantum Science and Technology, University of Calgary, Alberta T2N 1N4, Canada.
$^{8}$ Program in Quantum Information Science, Canadian Institute for Advanced Research, Toronto, Ontario M5G 1M1, Canada.
$^{9}$ Beijing Computational Science Research Center, Beijing 100084, China.\newline
$^*$Corresponding author, e-mail: junxiang@sxu.edu.cn
\end{affiliations}

\begin{abstract}
\baselineskip=0.8 cm We experimentally demonstrate coherent amplification of probe field in a tripod-type atoms driven by strong coupling, signal and weak probe fields.
We suppress linear and nonlinear atomic absorptions for resonant and near resonant probe via double electromagnetically induced transparency (DEIT).
Combining these advantages of suppressed absorption along
with temperature- or atomic-density-controlled transfer of population(ToP) between hyperfine ground states, we can induce near-resonant amplification of probe through stimulated Raman scattering(SRS) pumped by low-intensity signal field. The increased population difference of initial and final states of SRS due to increased ToP rate, together with reduced absorption at the second EIT window in an optically thick Cesium vapor, gives rise to highly effective coherent amplification.
\end{abstract}

\section*{Introduction}
Electromagnetically induced transparency (EIT), which arises from
destructive quantum interference between different transitions in a multi-level atom,
can exhibit total transparency even in the weak driving-field limit~\cite{Har97,FL00}.
Exploiting EIT quantum interference leads to applications
such as enhancing various linear and nonlinear optical processes~\cite{Har97,Harris90,Merriam2000,Hakuta1991}
including optical amplification~\cite{Imam89,Har89,Koc92}, which is
crucial not only for implementing quantum devices of all-optical switch, diode and transistors in optical network~\cite{HH00,DIC05,WZG13,JZL14,BDC13,CBB13,GTS14},
but also for investigating quantum resources with the exact frequency matched to the atomic resonant transition for long-term quantum communication~\cite{DLC01,BV05}.
In order to build integrated optical components of networks, e.g. a qubit amplifier based quantum repeater~\cite{GPS10} and EIT-based amplification in a waveguide~\cite{JBB10}, it is necessary to create the optical amplifiers operating in low noise, ultracompact and high efficiency~\cite{BPM13,JBB10}.

Amplification arises mainly via two mechanism: (i) suppressing resonant absorption via EIT while simultaneously incoherent pumping of atoms to the upper level
without population inversion~\cite{Har89,SF94,ZLN95,H95}; (ii) exploiting nonlinear processes such as frequency wave mixing~\cite{HH00,HMS92,LX96,HSH97}
or stimulated Raman scattering (SRS)~\cite{Cotter77,BML08,GZW14,Jin01,Jin02,Jin03}. The inversionless amplification with EIT-like spectral width showed the consistent evidence of coherent and resonant amplification~\cite{HH09,H08}, however this mechanism was not efficient enough to produce high gain.
Nonlinear amplification circumvents absorption-suppression and even allows the upper level to be unpopulated, but requires instead phase matching or a strong pump. Though EIT
was introduced in nonlinear schemes, the optimal high gain was
still obtained with high pump detuning, which is comparable to the Doppler width of
atomic transition to avoid the atomic absorption. Furthermore, the amplification was also proposed in double EIT(DEIT) system (induced by RF field coherent perturbation of single EIT) with the aid of incoherent pumping atoms to perturbation level~\cite{LYF99}, and the nonlinearity was demonstrated in this DEIT system via off-resonant RF field perturbation~\cite{YSK03}.
We combine both types of amplification in tripod DEIT atomic system in high density cesium vapor,
which enhances the efficiency and allows the amplifier to have near-resonant high gain with low pump power.
Specifically we demonstrate strong coherent amplification at one of DEIT. The DEITs  reduce not only both resonant and near-resonant absorptions, but also enhance the single effect of both of two EIT windows. As a result, the amplification of near resonant probe due to SRS is obtained, while the gain is highly enhanced via using high density atoms for increasing pumping rate of SRS. This tripod multilevel system coupling with multiple fields enables the coexistence of controllable DEIT or multiple EIT and double-DEIT windows for enhancing higher-order nonlinearity~\cite{ZBX07} or controlling group velocity~\cite{PK02}, and possible amplification~\cite{AS13}.

Using the second EIT window where the probe-field detuning equals signal-field detuning
but differs from coupling-field detuning is advantageous.
Through this window, we are able to attain highly reduced absorption due to enhanced EIT for resonant and near resonant probe,
and thus achieve near-resonant amplification pumped by signal field through two-photon-resonant SRS process, in which the Raman gain is proportional to the pump intensity of signal and population difference of initial and final states of Raman transition~\cite{Y.R.Shen}.
Here in our system, strong near-resonant amplification with low-intensity pump is possible because of the ability to populate the initial state of Raman transition within the ground-level hyperfine manifold by
our new mechanism we call transfer of population (ToP), which is mediated by spin-exchange collisions. In our case,
amplification depends not only on strong ground-state coherence between hyperfine levels as well as Zeeman sublevels
but also on the atomic density. Thus our high-gain amplification can be easily controlled via interacting fields and atomic temperature.

\section*{Results}
\subsection{Theory}
The tripod-type field-atom system is illustrated in the simplified energy level diagram of Fig.~1(a). It consists of an excited ($6^{2}P_{1/2},F_{e}=4$)
and two ground hyperfine states ($6^{2}S_{1/2}$, $F_{g}=3$,
$F_{g}=4$) of the $^{133}\text{Cs}$ D1 line with~$F$ the total angular momentum.
The atom is driven by strong coupling and signal fields at frequencies~$\omega_{\text{c}}$ and~$\omega_{\text{s}}$
on the $\pi$-transition of $F_{g}=3\leftrightarrow F_{e}=4$ and $F_{g}=4\leftrightarrow F_{e}=4$, respectively, while being probed by a weak field at $\omega_{\text{p}}$ on the $\sigma$-transition of
$F_{g}=4\leftrightarrow F_{e}=4$. The signal field acts as a pump to excite a Raman transition between the Zeeman ground states$\left|b_1\right\rangle$
and $\left|b_0\right\rangle$, generating Raman-shifted (anti-Stokes) radiation at the frequency of the probe field $\omega_\text{p}=\omega_\text{s}+\omega_{b_1b_0}$~\cite{Boyd},
$\hbar\omega_{b_1b_0}$ is the energy of the Raman transition.

\subsection{A. The population distribution in Zeeman sublevels.}
For such polarized-field-driven atoms, we consider the atomic structure with Zeeman sublevels denoted by $\left|c_i\right\rangle$, $\left|a_i\right\rangle$, and
$\left|b_i\right\rangle$ ($i=-m_{F},\cdots,m_{F}$ with $m_{F}$ the projection of $F$ along the quantization axis), as shown in Figs.~1(c,d). The strong coupling and signal fields are interacting with atoms on the $\pi$-transition of $\left|c_{i}\right\rangle\leftrightarrow\left|a_{i}\right\rangle$ and $\left|c_{i}\right\rangle\leftrightarrow\left|b_{i}\right\rangle$
with detuning $\delta_{\text{c}}=\omega_{\text{c}}-\omega_{c_{i}a_{i}}$ and  $\delta_{\text{s}}=\omega_{\text{s}}-\omega_{c_{i}b_{i}}$
($\omega_{c_{i}a_{i}}$ and $\omega_{c_{i}b_{i}}$ are the transition frequencies from $\left|c_{i}\right\rangle$ to $\left|a_{i}\right\rangle$
and $\left|b_{i}\right\rangle$) respectively. A weak field is applied to probe the $\sigma$-transitions of $\left|c_{i}\right\rangle\leftrightarrow\left|b_{i\pm1}\right\rangle$ with detuning
$\delta_{\text{p}}=\omega_{\text{p}}-\omega_{c_{i}b_{i\pm1}}$. Comparing with the strong coupling and signal pumping effects, we neglect the weak optical pumping effect of probe in this calculation for atomic population.
The optical Bloch equations of the system are as follows~\cite{GWR03}
\begin{eqnarray}
{\dot\rho_{c_{i}c_{i}} }&=&i\left(\rho_{c_{i}b_{i}}\Omega_{b_{i}c_{i}}-\Omega_{c_{i}b_{i}}\rho_{b_{i}c_{i}}
+\rho_{c_{i}a_{i}}\Omega_{a_{i}c_{i}}-\Omega_{c_{i}a_{i}}\rho_{a_{i}c_{i}}\right)
-\Gamma_{c_{i}}\rho_{c_{i}c_{i}}, \notag \\
{\dot\rho_{a_{i}a_{i}} }&=&i\left({\rho_{a_{i}c_{i}}\Omega_{c_{i}a_{i}}-\Omega_{a_{i}c_{i}}\rho_{c_{i}a_{i}}}\right)-\Gamma_{a_{i}}\rho_{a_{i}a_{i}}+(\dot\rho_{a_{i}a_{i}})_{\text{SE}}
+\gamma^{\prime}_{ba}\sum_{b_{j}}{\rho_{b_{j}b_{j}}}+\gamma^{\prime}_{aa}\sum_{a_{j},j\neq i}{\rho_{a_{j}a_{j}}},\notag \\
{\dot\rho_{b_{i}b_{i}} }&=&i\left({\rho_{b_{i}c_{i}}\Omega_{c_{i}b_{i}}-\Omega_{b_{i}c_{i}}\rho_{c_{i}b_{i}}}\right)-\Gamma_{b_{i}}\rho_{b_{i}b_{i}}+(\dot\rho_{b_{i}b_{i}})_{\text{SE}}
+{\gamma^{\prime}_{ab}\sum_{a_{j}}\rho_{a_{j}a_{j}}}+{\gamma^{\prime}_{bb}\sum_{b_{j},j\neq i}\rho_{b_{j}b_{j}}}, \notag \\
{\dot\rho_{a_{i}c_{i}} }&=&i\Omega_{a_{i}c_{i}}\left(\rho_{a_{i}a_{i}}-\rho_{c_{i}c_{i}}\right)
-\left(i\delta_\text{c}+\gamma_{c_{i}a_{i}}\right)\rho_{a_{i}c_{i}}, \notag \\
{\dot\rho_{b_{i}c_{i}} }&=&i\Omega_{b_{i}c_{i}}\left(\rho_{b_{i}b_{i}}-\rho_{c_{i}c_{i}}\right)-\left(i\delta_\text{s}+\gamma_{c_{i}b_{i}}\right)\rho_{b_{i}c_{i}}.
\label{eq1}
\end{eqnarray}
Where, $\Gamma_{k_i}$ is the decay rate for level $\left|k_{i}\right\rangle$, $\Gamma_{c_{i}}=4.6~\text{MHz}$ for Cs D1 line.
$\gamma_{k_{i}l_{j}}=\left(\Gamma_{k_i}+\Gamma_{l_j}\right)/2$ is the decoherence decay rate between levels $\left|k_{i}\right\rangle$ and $\left|l_{j}\right\rangle$ with $k,l\in\left\{a,b,c\right\}$. $\Omega_{c_{i}a_{j}}=-{\mu_{c_{i}a_{j}}E_\text{c}}/{2\hbar}$. ($\Omega_{c_{i}b_{j}}=-{\mu_{c_{i}b_{j}}E_\text{s}}/{2\hbar}$) is the effective Rabi frequencies for different Zeeman transitions $\left|c_{i}\right\rangle\leftrightarrow\left|b_{j}\right\rangle$ ($\left|a_{j}\right\rangle$) due to different Clebsch-Gordan coefficients~\cite{S09}.

It is noted that the ToP rate $\gamma^{\prime}_{ab}$
($\gamma^{\prime}_{ba}=9/7\gamma^{\prime}_{ab}$~\cite{LLX96}) between hyperfine ground levels of $\left|a\right\rangle$
and $\left|b\right\rangle$) in high density atoms is specifically concerned with the value of $\gamma^{\prime}_{ab}=\kappa N$ with $\kappa\approx6\times10^{-10}~\text{cm}^{3}~\text{s}^{-1}$~\cite{WW02,SFS08}. $\gamma^{\prime}_{ab}$ is dominated by spin-exchange collisions between atoms, and hence depends linearly on the atomic density
$N$.

On the contrary, however, the collisional decay rates $\gamma^{\prime}_{aa}$
($\gamma^{\prime}_{bb}$) between Zeeman sublevels are assumed to be small and constant as these rates are nearly
independent of the atomic density~\cite{SFS08}.
Neglecting collisional dephasing,
we have ${\Gamma_{a_i}}=9\gamma^{\prime}_{ab}+6\gamma^{\prime}_{aa}$, ${\Gamma_{b_i}}=7\gamma^{\prime}_{ba}+8\gamma^{\prime}_{bb}$.

Solving Eq.~(\ref{eq1}), we obtain the fractions of population in each Zeeman sublevel states, shown in Fig.~1(b), also shown by the numbers in Figs.~1(c,d). It is obviously seen that Population distribution can be modified by high ToP rate $\gamma^{\prime}_{ab}$.
For small ToP rate $\gamma^{\prime}_{ab}$=20~\text{Hz} with atomic density $N=5\times10^{10}~\text{cm}^{-3}$ in Fig.~1(c),
the whole population is in $\left|b_i\right\rangle$, especially in~$\left|b_0\right\rangle$.
However, a higher ToP rate $300~\text{Hz}$
employed for a higher density $N=5\times10^{11}~\text{cm}^{-3}$ in Fig.~1(d)
transfers population between $\left|b_i\right\rangle$ and $\left|a_j\right\rangle$,
giving rise to the decrease of population in $\left|b_0\right\rangle$, while the increase of that in other states
, especially in $\left|b_1\right\rangle$ (50 times the $\left|c_1\right\rangle$ population).
See also the Fig.~1(b) and its inset, the giant decrease of $\rho_{b_{0}b_{0}}$ and increase of $\rho_{b_{1}b_{1}}$ occurs when $\gamma^{\prime}_{ab}$ varies from $20~\text{Hz}$ to $600~\text{Hz}$ with corresponding atomic density $N=5\times10^{10}\sim1\times10^{12}\text{cm}^{-3}$ for
atomic temperature $25\sim80$ $^{\circ}\text{C}$~\cite{S09}.

\subsection{B. The transmission of the probe field.}
According to the population distribution in Figs.~1(c,d), the most atoms are populated at the states of $\left|c_i\right\rangle$, $\left|a_i\right\rangle$, and
$\left|b_i\right\rangle$ with $i=-1, 0, +1$. Therefore two efficient tripod structures are mainly constructed, $\left|b_{0}\right\rangle\leftrightarrow\left|c_{1}\right\rangle\leftrightarrow\left|a_{1}\right\rangle\leftrightarrow\left|b_{1}\right\rangle$
and $\left|b_{0}\right\rangle\leftrightarrow\left|c_{-1}\right\rangle\leftrightarrow\left|a_{-1}\right\rangle\leftrightarrow\left|b_{-1}\right\rangle$, and can be the simplified as a tripod model of Fig.~1(a) for further discussion.

With the increase of ToP rate via using high-density (or optically thick) atoms in a high-temperature vapor, the populations in $\left|a_1\right\rangle$, $\left|b_1\right\rangle$ and $\left|c_1\right\rangle$ are increased. Now we consider the case of Raman transition from $\left|b_1\right\rangle$ to $\left|b_0\right\rangle$, the signal-field pump atoms from $\left|b_1\right\rangle$ to $\left|c_1\right\rangle$ to stimulate emission from $\left|c_1\right\rangle$ to $\left|b_0\right\rangle$, resulting in the amplification for the probe. Therefore the increase of population in initial Raman state of $\left|b_1\right\rangle$ leads to an efficient SRS with enhancement of amplification. On the other hand, the partial increased population in upper level $\left|c_1\right\rangle$
can cause negligible stimulated emission without absorption in the EIT window.
The SRS can be obtained from probe absorption with relevant optical Bloch equations of coherence terms:
\begin{eqnarray}
{\dot\rho_{b_{0}c_{1}}}&=&i\left[\Omega_\text{p}\left(\rho_{b_{0}b_{0}}-\rho_{c_{1}c_{1}}\right)+\Omega_\text{c}\rho_{b_{0}a_{1}}+\Omega_\text{s}\rho_{b_{0}b_{1}}\right]
-\left(i\delta_\text{p}+\gamma_{c_1b_0}\right)\rho_{b_{0}c_{1}}, \notag \\
{\dot\rho_{b_{1}c_{1}}}&=&i\left[\Omega_\text{s}\left(\rho_{b_{1}b_{1}}-\rho_{c_{1}c_{1}}\right)+\Omega_\text{c}\rho_{b_{1}a_{1}}+\Omega_\text{p}\rho_{b_{1}b_{0}}\right]
-\left(i\delta_\text{s}+\gamma_{c_1b_1}\right)\rho_{b_{1}c_{1}}, \notag \\
{\dot\rho_{a_{1}c_{1}}}&=&i\left[\Omega_\text{c}\left(\rho_{a_{1}a_{1}}-\rho_{c_{1}c_{1}}\right)+\Omega_\text{s}\rho_{a_{1}b_{1}}+\Omega_\text{p}\rho_{a_{1}b_{0}}\right]
-\left(i\delta_\text{c}+\gamma_{c_1a_1}\right)\rho_{a_{1}c_{1}}, \notag \\
{\dot\rho_{b_{0}a_{1}}}&=&i\left(\rho_{b_{0}c_{1}}\Omega_\text{c}-\Omega_\text{p}\rho_{c_{1}a_{1}}\right)
-\left(i\delta_\text{pc}+\gamma_{b_0a_1}\right)\rho_{b_{0}a_{1}}, \notag \\
{\dot\rho_{b_{0}b_{1}}}&=&i\left(\rho_{b_{0}c_{1}}\Omega_\text{s}-\Omega_\text{p}\rho_{c_{1}b_{1}}\right)
-\left(i\delta_\text{ps}+\gamma_{b_1b_0}\right)\rho_{b_{0}b_{1}}, \notag \\
{\dot\rho_{b_{1}a_{1}}}&=&i\left(\rho_{b_{1}c_{1}}\Omega_\text{c}-\Omega_\text{s}\rho_{c_{1}a_{1}}\right)
-\left(i\delta_\text{sc}+\gamma_{b_1a_1}\right)\rho_{b_{1}a_{1}}.
\label{eq2}
\end{eqnarray}
Where, two-photon detuning $\delta_{ij}=\delta_{i}-\delta_{j}$ ($i,j=\text{c,s,p}$).
The state population terms are the values calculated in Section A. Hence, the probe absorption represented by the imaginary part of steady state density matrix element $\rho_{b_{0}c_{1}}$ is:
\begin{eqnarray}
	{\text{Im}\chi_\text{p}}
		&\propto&-\frac{A}{A^{2}+B^{2}}\left(\rho_{c_1c_1}-\rho_{b_0b_0}\right)\\ \nonumber
&-&\frac{\gamma_{b_1b_0}E-\delta_\text{ps}F}{\left(E^{2}+F^{2}\right)\left(\gamma_{b_1b_0}^{2}+\delta_\text{ps}^{2}\right)}\left[\left|\Omega_\text{p}\right|^{2}\left(\rho_{c_1c_1}-\rho_{b_0b_0}\right)
                                  -\left|\Omega_\text{s}\right|^{2}
                                  \left(\rho_{c_1c_1}-\rho_{b_1b_1}\right)\right].\\ \nonumber
	A&=&\gamma_{c_1b_0}+\frac{\left|\Omega_\text{c}\right|^{2}\gamma_{b_0a_1}}{\gamma_{b_0a_1}^{2}+\delta_\text{pc}^{2}}
+\frac{\left|\Omega_\text{s}\right|^{2}\gamma_{b_1b_0}}{\gamma_{b_1b_0}^{2}+\delta_\text{ps}^{2}},\;
	B=\delta_\text{p}-\frac{\left|\Omega_\text{c}\right|^{2}\delta_\text{pc}}{\gamma_{b_0a_1}^{2}+\delta_\text{pc}^{2}}
-\frac{\left|\Omega_\text{s}\right|^{2}\delta_\text{ps}}{\gamma_{b_1b_0}^{2}+\delta_\text{ps}^{2}},
\\\nonumber
	C&=&\gamma_{c_1b_1}+\frac{\left|\Omega_\text{c}\right|^{2}\gamma_{b_1a_1}}{\gamma_{b_1a_1}^{2}+\delta_\text{sc}^{2}}
+\frac{\left|\Omega_\text{p}\right|^{2}\gamma_{b_1b_0}}{\gamma_{b_1b_0}^{2}+\delta_\text{ps}^{2}},\;
	D=-\delta_\text{s}+\frac{\left|\Omega_\text{c}\right|^{2}\delta_\text{sc}}{\gamma_{b_1a_1}^{2}+\delta_\text{sc}^{2}}
-\frac{\left|\Omega_\text{p}\right|^{2}\delta_\text{ps}}{\gamma_{b_1b_0}^{2}+\delta_\text{ps}^{2}},\;
\\\nonumber
	E&=&CA-DB,\;
	F=CB+DA.
\label{eq3}
\end{eqnarray}
Where, $\gamma_{b_1b_0}=\gamma_{b_1a_1}=\gamma_{b_0a_1}=9\gamma^{\prime}_{ab}$, $\gamma_{c_1b_0}=\gamma_{c_1b_1}=\left(\Gamma_{c_1}+9\gamma^{\prime}_{ab}\right)/2$.
Equation~(3) describes several optical mechanisms:
Linear absorption with EIT effect for probe described by the first term of the right-hand side~(3), it can be neglected
under EIT condition for $\delta_\text{s}=\delta_\text{p}$ and for~\cite{AS14}
\begin{equation}
\left|\Omega_\text{s}\right|^2\gg\gamma_{b_1b_0}\gamma_{c_1b_0}.
\label{eq:Cond1}
\end{equation}

The second term represents the stimulated emission and nonlinear absorption, which is proportional to $\rho_{c_1c_1}-\rho_{b_0b_0}\left|\Omega_\text{p}\right|^2$, and implies the atom absorbs one probe photon to excited state, and down to ground state, and then absorbs a probe photon to excited state again, involving multi-wave process. Stimulated emission can be negligible since the probe intensity and population in $\left|c_1\right\rangle$ are small. Nonlinear absorption could also be eliminated by the third term under conditions of
\begin{equation}
	\rho_{b_1b_1} \left|\Omega_\text{s}\right|^2>\rho_{b_0b_0} \left|\Omega_\text{p}\right|^2.
\label{eq:Cond2}
\end{equation}

The probe amplification is represented by the third term, which is exactly the expression of SRS process pumped by signal field~\cite{Y.R.Shen}. This term of amplification is dominated when the two conditions~(\ref{eq:Cond1}) and~(\ref{eq:Cond2}) are satisfied simultaneously.

In order to get efficient SRS, both the linear and nonlinear absorption must be eliminated with the conditions~(\ref{eq:Cond1}) and~(\ref{eq:Cond2}) be satisfied. On the other hand, conditions~(\ref{eq:Cond1}) and~(\ref{eq:Cond2}) constrain in the lower value of the signal intensity. However the SRS is normally occurred with strong pump, here, the pump is the signal field.

In our designed tripod-type system with high ToP rate, the higher population in $\left|b_1\right\rangle$ induced by high ToP rate increases the pump rate and leads to greater reduction in the signal intensity to attain Raman gain. High gain thus occurs using a low intensity signal field.

For Doppler-broadened system, the driven thermal atoms experience Doppler shifts leading to the version of Eq.~(3) averaged over the Maxwell velocity distribution $f\left(v\right)=\text{exp}\left(-v^{2}/u^{2}\right)/u\sqrt{\pi}$.
$k$ is the input laser wavenumber, $v$ is the atomic velocity along the field propagation direction,
$u=\sqrt{2k_\text{B}T_\text{at}/M}$ is the most probable atomic velocity ($k_\text{B}$ is the Boltzmann's constant, $T_\text{at}$ is atom temperature, and $M$ is the atomic mass).
The condition~(\ref{eq:Cond1}) is modified to~\cite{AS14}
\begin{equation}
	\left|\Omega_\text{s}\right|^2\gg\gamma_{b_1b_0}(\gamma_{c_1b_0}+W_\text{D}),
\label{eq:Cond3}
\end{equation}
with~$W_\text{D}$ the Doppler-broadening width.

Considering the condition of Eq.~(\ref{eq:Cond3}), we notice that the higher~$W_\text{D}$ in high-temperature atoms lets the Eq.~(\ref{eq:Cond3}) be difficultly satisfied with low-intensity signal field. However, in the following, we show that the strong amplification due to SRS is achieved in high-temperature atomic cell. Rather, the amplification can not be found in room-temperature atomic cell.
We theoretically and experimentally show good agreement that the high efficient Raman amplification with low signal pump is obtained with the aid of increased ToP rates in high-temperature atomic cell.

Figures~2(a,b) show the probe transmission spectrum expressed by $T=\exp\left[-kL\text{Im}\chi_\text{p}\left(\delta_\text{p}\right)\right]$
through a length-$L$ atoms
with two different ToP rates in different densities or optical depth of atoms.

As shown in Fig.~2(a) in low optical depth of atoms, two EIT windows is observed. One window is obtained at one- and two-photon resonances $\delta_\text{p}=0$ and $\delta_\text{pc}=0$ for the coupling and probe fields,
which forms a typical $\Lambda$-type EIT when the signal field is absent.
The second EIT window is found at single-photon detuning
but at two-photon resonance ($\delta_\text{p}=10~\text{MHz}$,
$\delta_\text{ps}=0$) for the signal and probe fields.

Surprisingly, we note here that an amplification in the second EIT window is obtained in Fig.~2(b) when a large TOP rate is taken into account in high-density atoms. Although the signal intensity suffices to satisfy condition~(\ref{eq:Cond3}) for the cases shown in Fig.~2, amplification does not exist in Fig.~2(a) until condition~(\ref{eq:Cond2}) is satisfied as shown in Fig.~2(b). The large ToP rate causes a higher population in state $\left|b_1\right\rangle$ reaching $\rho_{b_1b_1}\approx0.051$, which is high enough to satisfy condition (\ref{eq:Cond2}) for the case of amplification shown in Fig.~2(b).

We also plot the probe transmission in the theoretical absence of the third term for SRS (see the blue lines in Fig.~2).
Evidently amplification vanishes for both cases of small and of large ToP rates.
Confirming our earlier assertion,
amplification is a result of both absorption suppression at the second EIT window contributed by the first and second terms of Eq.~(3) and SRS process enhanced by ToP rate, which leads to a population increase of~$\left|b_1\right\rangle$ and of~$\left|c_1\right\rangle$.
Consequently, increasing population in~$\left|b_1\right\rangle$ will make the condition~(\ref{eq:Cond2}) be satisfied at low pump intensity of signal field, thereby enhancing the SRS pump rate to obtain high gain.

\subsection{Experiment}

The experimental implementation of obvious coherent induced amplification is shown in Fig.~3. We perform the experiment
in a 75-$\text{mm}$-long $^{133}\text{Cs}$ vapor cell with anti-Reflection (AR) coated
end windows. The cell is wrapped with three layers of $\mu$-metal
cylinder to shield the stray magnetic field.
The coupling, signal and probe lights are from separated diode laser sources,
driving the atomic transitions (as shown in Fig.~1) and co-propagating through
the vapor cell with a small angle about $7~\text{mrad}$.
The vertically linearly polarized coupling and signal lights are all
frequency stabilized using their own interlock controller systems.
The horizontally linearly polarized probe light is frequency scanned over
the atomic resonance to measure the absorption spectrum as well as the Doppler-broadened absorption profile. The coupling, signal and probe powers are
$P_\text{c}=30~\text{mW}$, $P_\text{s}=10~\text{mW}$ and $P_\text{p}=20~\mu\text{W}$, corresponding to the Rabi frequencies $\Omega=\alpha\Gamma_{c_1}\sqrt{(P/\pi r^{2})/2I_\text{sat}}$
of $23~\text{MHz}$, $7~\text{MHz}$ and $1~\text{MHz}$ for the beam radius $r$ of
$1~\text{mm}$, $0.5~\text{mm}$, and $0.5~\text{mm}$, respectively
($\alpha$ is C-G coefficient, $I_\text{sat}=2.5~\text{mW/cm}^{2}$ is saturated intensity~\cite{S09}). The probe absorption is measured via observing the probe transmission spectra, which is inversely proportional to exponential function of absorption, and evaluated as the value of transmission intensity of
probe over the input intensity at far-detuned frequency. The transmission $T$ with the value smaller than 1 shows the absorption property, while the larger than 1 of $T$ means the amplification gain $g$.

We show probe transmission
vs probe detuning in Figs.~2(c,d).
We see two EIT peaks within the Doppler-broadened background dip in Fig.~2(c) under two-photon
resonance conditions $\delta_\text{p}=\delta_\text{c}=0$ for hyperfine coherence and
$\delta_\text{p}=\delta_\text{s}=10~\text{MHz}$ for Zeeman coherence in a room-temperature vapor cell with optical depth $\text{OD}\approx1.5$.
Compared with typical EIT in a $\Lambda$ system without signal as shown in Fig.~2(c) (also shown in the inset),
EIT enhancement for two-photon resonance $\delta_\text{p}=\delta_\text{c}=0$ is readily apparent due to the second transparency. It also demonstrates the enhancement of Zeeman coherence of EIT, the linewidth of which is narrower than that of EIT for only applied signal and probe fields in an open degenerate two-level system, in which the broad background transmission EIT peak is introduced by both the optical pumping and saturation effects~\cite{KMK03,KMH01}.The theoretical results In Fig 2(b) show the transmission of the first EIT is nearly 1, while the experiment has only <0.5 in Fig. 2(d). The difference occurs mainly due to the linewidth of driving fields, atomic decoherence and diffusion, which is not considered in theoretical model~\cite{EITX01}. It is noted that the Zeeman EIT has lower OD, because it is observed in an open degenerate two-level system, in which the broad background transmission EIT peak surrounding an electromagnetically induced absorption (EIA) effect is introduced by both the optical pumping and saturation effects.

We increase the temperature of the vapor cell to
$T_\text{at}=65^{\circ}\text{C}$
with high atomic density or higher $\text{OD}\approx18$, which supports high ToP rates. As shown in Fig.~2(d), an amplification peak at the second window of EITs within highly absorbed Doppler-broadened background
is then obtained. It is also shown in the inset of Fig.~2(d) that the spectral width of the gain is 1.7MHz, and is narrower than the spectral width of EIT due to highly nonlinear process with strong coherence. The spectral width depends mainly on the quantum coherence of atoms and nonlinear effects determined by the Raman pump intensity and Raman detuning. This gain observation supports our theoretical prediction that higher ToP rates in higher temperature atoms lead to amplification due to increasing population in $\left|b_1\right\rangle$, which satisfies the condition~(\ref{eq:Cond2}) for amplification. Compared with the case of room temperature atoms in Fig.~2(c), in which amplification does not occur i.e., the condition~(\ref{eq:Cond2}) is not reached, the larger Doppler width $W_\text{D}$ in higher temperature atoms makes the condition even be difficult to reach unless the population in $\left|b_1\right\rangle$ is highly increased.

Now we show that increasing temperature can increase the ToP rate, thereby increasing population in $\left|b_1\right\rangle$ to get higher probe gain. Fig.~4(a) shows the increase of gain with increasing atom temperature in the region of $60\sim70^{\circ}\text{C}$, surrporting the effect of ToP rates on the amplification. The experimental data show good agreement with the theoretical prediction in the solid line for $60\sim70^{\circ}\text{C}$. Further increase of temperature from 70 to 80$^{\circ}\text{C}$ reduces the gain. The experimental results over $70^{\circ}\text{C}$ can not fit the theoretical calculation, since in higher dense atoms, ToP leads to the nonlinear term of Eq.~(\ref{eq:Cond2}) be easy satisfied while it weaken the EIT condition of Eq.~(\ref{eq:Cond3}) due to increasing of $\gamma_{b_{1}b_{0}}$ and $W_\text{D}$, the values of which may be inexactly estimated, and the other effects e.g., the inhomogeneous dephasing and collisional loss in dense atoms are also neglected in theory. We can also increase the Rabi frequency of pumping-signal field to overcome this trade-off effect, see the black squares in Fig.~4(a).

The amplification is induced by quantum coherences and SRS precess. The efficiency or gain depends not only on the ToP rate but also on quantum coherences, e.g., Zeeman coherence
between ground levels $\left|b_0\right\rangle$ and $\left|b_1\right\rangle$, which is mainly determined by experimental parameters, such as spontaneous decay of the upper level, collisions between the atoms, and the phase and frequency stability of the laser fields, and can be improved experimentally by using
phase-locked signal and probe fields from a laser beam split by a beam splitter~\cite{MS08}. In Fig.~4(b), we compare the probe gain vs the signal detuning using independent (black) and phase-locked (blue) signal and probe, which is scanned 60~$\text{MHz}$
by double passing two acousto-optical modulators during the measurement.
As shown in Fig.~4(b), the gain is significantly improved by using phase-locked laser system, e.g., the maximum gain $g=20$ at $\delta_\text{p}=\delta_\text{s}=10~\text{MHz}$, which is three times higher than $g=6.5$ obtained with the independent laser fields,
thereby showing the important contribution of Zeeman coherence on gain experimentally.

Moreover, Fig.~5(a) shows increased gain as signal power increases as expected from the third term in Eq.~(3) and Eq.~(\ref{eq:Cond2}) for SRS process. Increasing signal power could yield more gain,
but, in practice, the maximum diode-laser output power limits the ability to achieve high signal power. Gain also increases
with the increased coupling power, as shown in Fig.~5(b), coupling  power as low as $5~\text{mW}$
(equivalent to $\Omega_\text{c}=8~\text{MHz}$ ensures an EIT condition)
enables observation of gain, but a power as high as $30~\text{mW}$ optimizes gain in our system. The pump depletion effect stops the gain from rising
when $P_\text{c}>30~\text{mW}$. This amplification scheme does not need an intense pumping
field for high-gain observation. Figure~5(b) shows that high gain of $g=80$ can be obtained with only 30~mW coupling and signal fields.

\section*{Discussion}

In summary, by combining quantum coherence and SRS in multi-field-coupled atomic system with high spin-exchange collision-induced ToP rates, the controllable high-gain amplification near an atomic resonance was demonstrated with low pump power. This study has potential applications in all-optical components of transistor, diode, switch and nonclassical resource for quantum communication network.

\section*{Acknowledgements}

This work was supported by National Natural Science Foundation of China (11574188;GG2340000241), the Project for Excellent
Research Team of the National Natural Science Foundation of China (61121064),
the China 1000 Talent Plan, and AITF.

\section*{Author contributions}
J. Zhang conceived the original idea. D. Wang and J. Zhang
designed the experiment. D. Wang, C. Liu and C. Xiao constructed and
performed the experiment. D. Wang and J. Zhang accomplished theoretical
calculation and the data analyses. D. Wang and J. Zhang wrote the paper.
J. Zhang, B. C. Sanders,  H. M. M. Alotaibi, and L. Wang improved the writing.
B. C. Sanders, H. M. M. Alotaibi, L. Wang and S. Zhu provided helpful discussion.

\section*{Additional information}

Competing financial interests: The authors declare no competing financial
interests.

\newpage
\textbf{Figure 1:}
(a) The simplified field-atom system. Green, blue and black lines designate the coupling, pumping-signal and weak probe fields, respectively.
(b) Population $\rho_{kk}$ vs $\gamma^{\prime}_{ab}$. Dash-dot (pink), dash (blue), solid (red) and dot (black) curves
for $\rho_{b_0b_0}$, $\rho_{b_1b_1}$, $\rho_{a_1a_1}$ and $\rho_{c_1c_1}$, respectively.
The inset is the expanded view of $\rho_{b_1b_1}$, $\rho_{a_1a_1}$ and $\rho_{c_1c_1}$.
(c,d) The Zeeman sublevels with ToP rate $\gamma^{\prime}_{ab}$: $20~\text{Hz}$ (c) and $300~\text{Hz}$ (d).
Other parameters are: $\Gamma_{c_1}=4.6~\text{MHz}$, $\Omega_\text{c}=20~\text{MHz}$, $\Omega_\text{s}=8~\text{MHz}$,
$\delta_\text{c}=0$, $\delta_\text{s}=10~\text{MHz}$, $\gamma^{\prime}_{aa}=\gamma^{\prime}_{bb}=20~\text{Hz}$.

\textbf{Figure 2:} \textbf{The probe transmission spectra.} (a,b) for theoretical simulation with (a) $N=5\times10^{10}~\text{cm}^{-3}$, $\gamma^{\prime}_{ab}=20~\text{Hz}$, $\gamma_{b_1b_0}=340~\text{Hz}$, (b) $N=5\times10^{11}~\text{cm}^{-3}$, $\gamma^{\prime}_{ab}=300~\text{Hz}$, $\gamma_{b_1b_0}=2.86~\text{kHz}$.
Other parameters are the same as those in Fig.~1;
(c,d) for experiment at (c) room temperature, (d) $T_\text{at}=65^{\circ}\text{C}$.
Red, gray and black curves for DEIT, hyperfine coherence of EIT when the signal is absent and
Zeeman coherence of EIT without the coupling.

\textbf{Figure 3:} \textbf{Experimental sketch.} The weak uniform magnetic field B along
the $\textit{y}$-direction is applied to define the quantization axis. PBS: polarizing beam splitter, PD: photodetector.
$\updownarrow$ denotes the linear polarization.

\textbf{Figure 4:}
Gain at $\delta_\text{ps}=0$ vs (a)~$T_\text{at}$ for different $P_{\text{s}}$: 10~mW (blue circles), 20~mW (black squares); and (b)~$\delta_\text{s}$ with phase-locked (blue) and independent (black) signal and probe fields.
(a) The atomic density is $4\sim10\times10^{11}~\text{cm}^{-3}$ with ToP rate $240\sim600~\text{Hz}$, $\delta_\text{p}=10~\text{MHz}$; (b)~$T_\text{at}=65^{\circ}\text{C}$. Other parameters are the same as those in Figs.~2(b,d). The solid line represents the theory.

\textbf{Figure 5:}
(a) Gain vs $P_\text{s}$ for $P_\text{c}=30~\text{mW}$; (b) Gain vs $P_\text{c}$ for different $P_\text{s}$: 30~$\text{mW}$ (blue circles), 20~$\text{mW}$ (black triangles), and 10~$\text{mW}$ (pink squares). Other parameters are the same as those in Fig.~4(b). Solid curve is the theoretical predication.


\begin{figure}
\centerline{
\includegraphics[width = 15cm]{fig1.eps}
} \label{fig1}
\end{figure}

\begin{figure}
\centerline{
\includegraphics[width = 14cm]{fig2.eps}
} \label{fig2}
\end{figure}

\begin{figure}
\centerline{
\includegraphics[width = 9cm]{fig3.eps}
} \label{fig3}
\end{figure}

\begin{figure}
\centerline{
\includegraphics[width = 12cm]{fig4.eps}
}  \label{fig4}
\end{figure}

\begin{figure}
\centerline{
\includegraphics[width = 12cm]{fig5.eps}
} \label{fig5}
\end{figure}

\end{document}